\documentclass[11pt]{article}
\usepackage{amsfonts}
\usepackage{amssymb}
\usepackage{amsmath}
\usepackage[doublespacing]{setspace}
\usepackage{float}

\def\sqr#1#2{{\vcenter{\vbox{\hrule height.#2pt
        \hbox{\vrule width.#2pt height#1pt \kern#2pt
        \vrule width.#2pt}
        \hrule height.#2pt}}}}

\newcommand{\nc}{\newcommand}
\nc{\parent}[1]{$[\![#1]\!]$}
\newtheorem{theorem}{Theorem}[section]
\newtheorem{lemma}{Lemma}[section]

\newtheorem{proposition}{Proposition}[section]
\newtheorem{remark}{Remark}[section]
\newtheorem{definition}{Definition}[section]
\newtheorem{assumption}{Assumption}[section]

\newenvironment{proof}{{\sc Proof.}\hspace{3mm}}{\qed}
\newenvironment{pf-main}{{\sc Proof of Theorem \ref{mainresult}.}\hspace{3mm}}{\qed}
\nc{\cadlag}{c\`{a}dl\`{a}g } \nc{\ba}{\begin{array}}
\nc{\ea}{\end{array}} \nc{\be}{\begin{equation}}
\nc{\ee}{\end{equation}} \nc{\bea}{\begin{eqnarray}}
\nc{\eea}{\end{eqnarray}} \nc{\bean}{\begin{eqnarray*}}
\nc{\eean}{\end{eqnarray*}} \nc{\bu}{\bullet} \nc{\nn}{\nonumber}
\nc{\cA}{{\mathcal A}} \nc{\cB}{{\mathcal B}} \nc{\cC}{{\mathcal
C}} \nc{\cD}{{\mathcal D}} \nc{\bbD}{\mathbb{D}}
\nc{\cG}{{\mathcal G}} \nc{\cF}{{\mathcal F}} \nc{\cS}{{\mathcal
S}} \nc{\cU}{{\mathcal U}} \nc{\cH}{{\mathcal H}}
\nc{\cK}{{\mathcal K}} \nc{\cM}{{\mathcal M}} \nc{\cO}{{\mathcal
O}} \nc{\cP}{{\mathcal P}} \nc{\bbE}{\mathbb{E}}
\nc{\bbEQ}{\mathbb{E}_{\mathbb{Q}}} \nc{\eps}{\varepsilon}
\nc{\bbEP}{\mathbb{E}_{\mathbb{P}}}\nc{\bbL}{\mathbb{L}}
\nc{\bbP}{\mathbb{P}} \nc{\bbQ}{\mathbb{Q}} \nc{\del}{\partial}
\nc{\Om}{\Omega} \nc{\om}{\omega} \nc{\bbR}{\mathbb{R}}
\nc{\bbC}{\mathbb{C}} \nc{\bfr}{\begin{flushright}}
\nc{\efr}{\end{flushright}} \nc{\dXt}{\Delta X_{t}}
\nc{\dXs}{\Delta X_{s}} \nc{\bs}{\blacksquare} \nc{\dX}{\Delta X}
\nc{\dY}{\Delta Y}
\nc{\dnkx}{\left(X(T^{n}_{k})-X(T^{n}_{k-1})\right)}
\nc{\esssup}{\mathrm{ess}\mbox{ }\mathrm{sup}}
\nc{\essinf}{\mathrm{ess}\mbox{ } \mathrm{inf}}
\nc{\dhats}{\widehat{\delta_s}}

\nc{\qed}{\hfill $\blacksquare$}

\nc{\chf}{\mbox{$\mathbf1$}}
\begin{document}

\title{Stock Market Insider Trading in Continuous Time with Imperfect
Dynamic Information \thanks{
I benefited from helpful comments from Peter Bank, Rene Carmona, Christian Julliard, Dmitry Kramkov, Michael Monoyios, Andrew Ng, Bernt \O ksendal and seminar and workshop participants at 4th Oxford - Princeton Workshop, 14th Mathematics and Economics Workshop -- University of Oslo, Stochastic Filtering and Control Workshop -- Warwick University and Warwick Business School.}}
\author{Albina Danilova\thanks{%
Department of Mathematical Sciences, Carnegie Mellon University, Pittsburgh, PA 15213-3890, USA, e-mail: danilova@andrew.cmu.edu} \\
\emph{Department of Mathematical Sciences}\\
\emph{Carnegie Mellon University}}
\maketitle

\begin{abstract}
This paper studies the equilibrium pricing of asset shares in the
presence of \emph{dynamic} private information. The market consists
of a risk-neutral informed agent who observes the firm value, noise
traders, and competitive market makers who set share prices using
the total order flow as a noisy signal of the insider's information.
I provide a characterization of all optimal strategies, and prove
existence of both Markovian and non Markovian equilibria by deriving
closed form solutions for the optimal order process of the informed
trader and the optimal pricing rule of the market maker. The
consideration of non Markovian equilibrium is relevant since the market
maker might decide to re-weight past information after receiving a
new signal. Also, I show that $a)$ there is a unique Markovian
equilibrium price process which allows the insider to trade
undetected, and that $b)$ the presence of an insider increases the
market informational efficiency, in particular for times close to
dividend payment.
\end{abstract}

\newpage

\section{Introduction}

Although financial markets with informational asymmetries have been widely
discussed in the market microstructure literature (see \cite{Br} and \cite{Oh} for a review), the characterization of the optimal
trading strategy of an investor who posses superior information has been,
until lately, largely unaddressed by the mathematical finance literature.

In recent years, with the development of enlargement of filtrations
theory (see \cite{Ma}), models of so called insider
trading have been gaining attention in mathematical finance as well
(see e.g. \cite{An}, \cite{Bi} and \cite{Ka}). The salient assumptions of these
models are that $i)$ the informational advantage of the insider is a
functional of the stock price process (e.g. the insider might know
in advance the maximum value the stock price will achieve), and that
$ii)$ the insider does not affect the stock price dynamics. But in
fact, since equilibrium stock prices should clear the market, and
thus depend on the future random demand of market participants,
assuming that the informational advantage of the insider is a
functional of the price process implies that she either knows the
future demand processes of \textit{all market participants,} or she
knows that the price will -- exogenously -- converge to a
fundamental that is known to her. Since the assumption of an
omniscient insider is unrealistic, one would have to assume the
latter. Nevertheless, since the presence of an insider -- by
assumption in these models-- does not affect the price process, this
raises the question of what makes the price converge to its
fundamental value without information being released to the market.

Thus, from the market microstructure point of view, these modeling
assumptions translate into $i)$ imposing strong efficiency of the markets
even without an insider providing, through her trading, information to the
market -- that is, assuming a priori that the price will converge to the
fundamental value -- and that $ii)$ the less informed agents are \emph{not}
fully rational, since they do not try to infer the insider's private signal
from market data (since there is no feedback from insider trading to
equilibrium price).

Part of the mathematical finance literature has tried to address these
shortcomings by considering the informational content of stock prices, and
optimal information-based trading, in a rational expectations equilibrium
framework (see e.g. \cite{B}, \cite{Ch}). In these models -- to preserve
tractability -- the private information of the insider has been generally
assumed to be \emph{static}. For example, in \cite{B} and in \cite{Ch}
the insider knows \emph{ex ante }the final value of the firm, and in \cite{Ca} she knows \emph{ex ante} the time of default of the company
issuing the asset. This literature has shown that $i)$ the presence of an
insider on the market does not necessarily lead to arbitrage (i.e. the value
function of the insider is finite), and that $ii)$ the presence of insiders
might be considered beneficial to the market, in the sense that it leads to
higher information efficiency of the equilibrium price process.

Nevertheless, the assumption of insider's perfect foresight is unrealistic,
since the fundamental value of the firm should be connected to elements
(like future cash-flows, productivity, sales etc.) that have intrinsically
an aleatory component. That is, a more natural assumption would be that the
fundamental value is in itself a stochastic process, and that the insider
can observe it directly -- or at least observe it in a less noisy way than
the other agents on the market.

Thus, in this paper I relax the assumption of static insider
information, and study the equilibrium trading and price processes,
as well as market efficiency, in a setting with \emph{dynamic}
private information.

The model I consider in this paper is a generalization of the static
information setting of \cite{B}. An earlier attempt to generalize
this framework to include dynamic information is in \cite{BP}. This latter paper considers a much smaller set of
admissible trading strategies and pricing rules, and has much more
stringent assumptions on the parameters, than the ones considered in
my work. Moreover, it shows the existence of \emph{one} possible
Markovian equilibrium, while my work characterizes \emph{all
}optimal strategies and establishes that there is a unique
\emph{Markovian inconspicuous} equilibrium price process, i.e. an
equilibrium price that allows the insider to trade undetected and
depends only on the total order process. Moreover, I identify this
Markovian equilibrium in closed form, and show that the presence of
an insider increases the market informational efficiency for times
close to dividend payment. Furthermore, I show that even when the
market parameters do not satisfy the conditions for the existence of
a Markovian equilibrium, there exists a \emph{non Markovian}
inconspicuous equilibrium which I also identify in closed form.
Additionally, I give characterization of all optimal trading
strategies for the equilibrium price process. I show, based on this
characterization, that in the case of non Markovian price process it
is optimal for the insider to reveal her private information not
only at the terminal time, but also at some predefined interim times --
thus bringing the market to higher efficiency than in the case of
Markovian price process.

The remainder of the paper is organized as follows. Section 2
presents the model and the assumptions. Existence of Markovian
equilibrium, and uniqueness of the inconspicuous Markovian
equilibrium price process, are proved in Section 3. Existence of
equilibrium for more general pricing functionals is demonstrated in
the Section 4. Section 5 concludes.

\section{The Model Setup}

Consider a stock issued by a company with fundamental value given by the
process $Z_{t}$, defined on $\left( \Omega ,\mathcal{F},\left( \mathcal{F}%
_{t}\right) _{t\geq 0},\mathbb{P}\right) $, and satisfying
\begin{equation*}
Z_{t}=v+\int_{0}^{t}\sigma _{z}(s)dB_{s}^{1}
\end{equation*}%
where $B_{t}^{1}$ is a standard Brownian motion on $\mathcal{F}_{t}$, $v$ is
$N(0,\sigma )$ independent of $\mathcal{F}_{t}^{B^{1}}$ for any $t$, and $%
\sigma _{z}(s)$ a is deterministic function.

Then, if the firm value is observable, the fair stock price should be a
function of $Z_{t}$ and $t$. However, the assumption of the company value being
discernable by the whole market in continuous time is counterfactual, and it
will be more realistic to assume that this information is revealed to the
market only at given time intervals (such as dividend payments times or when
balance sheets are publicized).

In this model I therefore assume, without loss of generality, that
the time of the next information release is $t=1$, and the market
terminates after that.\footnote{This is without loss of generality,
since the extension to multiple information release times is
straightforward.} Hence, in this setting the stock can be viewed as
a European option on the firm value with maturity  $T=1$ and payoff
$f(Z_{1})$. In addition to this risky asset, there is a riskless
asset that yields an interest rate normalized to zero for
simplicity of exposition. In what follows it is assumed that all
random
variables are defined on the same stochastic basis $\left( \Omega ,\mathcal{F%
},\left( \mathcal{F}_{t}\right) _{t\geq 0},\mathbb{P}\right) $.

The microstructure of the market, and the interaction of market
participants, is modeled as a generalization of \cite{B}. There are
three types of agents: noisy/liquidity traders, an informed
trader (insider), and competitive market makers, all of whom are risk
neutral. The agents differ in their information sets, and objectives, as
follows.

\begin{itemize}
\item \textit{Noisy/liquidity traders} trade for liquidity reasons, and
their total demand at time $t$ is given by a standard Brownian motion $%
B_{t}^{2}$ independent of $B^{1}$ and $v$.

\item \textit{Market makers} observe only the total market order process $%
Y_{t}=\theta _{t}+B_{t}^{2}$, where $\theta _{t}$ is the total order of the
insider, i.e. their filtration is $\mathcal{F}_{t}^{M}=\bar{\mathcal{F}}%
_{t}^{Y}$. Since they are competitive and risk neutral, on the basis
of the observed information they set the price as
\begin{equation}
P\left( Y_{[0,t]},t\right) =P_{t}=\mathbb{E}\left[ f(Z_{1})|\mathcal{F}%
_{t}^{M}\right] \text{.}  \label{mm_obj}
\end{equation}%
As in \cite{Ch}, I assume that market makers set the price as a
function of weighted total order process at time $t$, i.e. I
consider pricing functionals $
P\left( Y_{[0,t]},t\right) $ of the following form%
\begin{equation*}
P\left( Y_{[0,t]},t\right) =H\left(\int_0^tw(s)dY_s,t\right).
\end{equation*}
where $w(s)$ is some positive deterministic function.
\item \textit{The informed investor} observes the price process $
P_{t}=H(\int_{0}^tw(s)dY_s,t) $ and the true firm value $Z_{t}$,
i.e. her filtration is given by
$\mathcal{F}_{t}^{I}=\bar{\mathcal{F}}_{t}^{Z,P}$. Since she is
risk-neutral, her objective is to maximize the expected final
wealth, i.e.
\begin{equation}
\sup_{\theta \in \mathcal{A}(H,w)}\mathbb{E}\left[ X_{1}^{\theta
}\right] =\sup_{\theta \in \mathcal{A}(H,w)}\mathbb{E}\left[
(f(Z_{1})-P_{1})\theta _{1}+\int_{0}^{1}\theta _{s-}dP_{s}\right]
\label{ins_obj}
\end{equation}%
where $\mathcal{A}(H,w)$ is the set of admissible trading strategies
for the given price functional $H\left( \int_0^{t}w(s)dY_s,t\right)
$ which will be defined later. That is, the insider maximizes the
expected value of her final wealth
$X_{1}^{\theta }$, where the first term on the right hand side of equation (%
\ref{ins_obj}) is the contribution to the final wealth due to a potential
differential between price and fundamental at the time of information
release, and the second term is the contribution to final wealth coming from
the trading activity.
\end{itemize}

Note that setting $\sigma _{z}\equiv 0$, the resulting market would
be the static information one considered by \cite{B}.

Note also that the above market structure implies that the insider's
optimal trading strategy takes into account the \emph{feedback
effect }i.e. the that prices react to her trading strategy according
to equation (\ref{mm_obj}). Identifying the optimal insider's
strategy is equivalent to the problem of finding the rational
expectations equilibrium of this market, i.e. a pair consisting of
an \emph{admissible} price functional and an \emph{admissible}
trading strategy such that: \textit{a)} given the price functional
the trading strategy is optimal, and \textit{b)} given the trading
strategy the price functional satisfies (\ref{mm_obj}). To formalize
this definition, we first need to define the sets of admissible
pricing rules and trading strategies.

Although it is standard in the insider trading literature to limit
the set of admissible strategies to absolutely continuous ones, in what
follows I consider a much broader class of strategies given by the
set of semimartingales satisfying some standard technical conditions
that eliminate doubling strategies. The formal definition of the set
of admissible trading strategies is summarized in the following
definition.

\begin{definition}
An insider's trading strategy, $\theta _{t}$, is admissible for a
given
pricing rule $(H(y,t),w(t))$ ($\theta \in \mathcal{A}(H,w)$) if $\theta _{t}$ is $%
\mathcal{F}_{t}^{I}$ adapted semimartingale, and no doubling strategies are
allowed i.e.%
\begin{equation}
\mathbb{E}\left[ \int_{0}^{1}H^{2}\left(\int_0^tw(s)d\theta
_{s-}+\int_0^tw(s)dB_{s}^{2},t\right)dt\right] <\infty .
\label{theta_cond_2}
\end{equation}%
Moreover, we call the insider's trading strategy inconspicuous if $%
Y_{t}=\theta _{t}+B_{t}^{2}$ is a Brownian motion on its own filtration $%
\mathcal{F}_{t}^{Y}$ (since in this case the presence of the insider is
undetectable).
\end{definition}

\begin{remark}
An equilibrium in which the optimal insider's trading strategy is
inconspicuous is a desirable feature of any insider trading model, and I
will show that in this setting such an equilibrium exists. In
fact, given the potentially high cost associated with being identified as an
insider, it might be reasonable to consider only this type of equilibrium.
\end{remark}

The definition of admissible pricing rules is a generalization
 of the one in \cite{B}{\footnote{Setting $w(s)\equiv 1$ will make conditions 2-4 exactly
the same as in \cite{B}}} with additional regularity condition
\textit{5} below which insures that, given the market maker's filtration,
the total order process has finite variance. This generalization
allows the market maker to re-weight her past information.

\begin{definition}
\label{pr_rule_def} A pair of measurable functions, $H\in
C^{2,1}(\mathbb{R}\times \lbrack 0,1])$, $H:\mathbb{R}\times \lbrack
0,1]\rightarrow \mathbb{R}$ and $w:[0,1]\rightarrow
\mathbb{R}_{+}\backslash \{ 0 \}$, is an admissible pricing rule
($(H,w)\in \mathcal{H}$) if and only if:

\begin{enumerate}
\item  The weighting function, $w(t)$, is a piecewise positive constant function
given by
\begin{equation}\label{def:weight_func}
w(t)=\sum_{i=1}^{n}\sigma^i_y 1_{\{t\in (t_{i-1},t_i]\}}
\end{equation}
where $0=t_0<t_1< \ldots <t_n=1$ and $\sum
_{i=1}^{n}(\sigma^i_y)^2=1$.

This condition doesn't cause loss of generality because: \emph{a)}
it was shown by \cite{Ch}, in the static private information case,
that in the equilibrium $w'(t)= 0$ and \emph{b)} it is always
possible to re-scale $w$ to have $\sum _{i=1}^{n}(\sigma^i_y)^2=1$.
\item $\mathbb{E}\left[ \int_{0}^{1}H^{2}(\int_0^tw(s)dB_{s}^{2},t)dt\right] <\infty .$

\item $\mathbb{E}\left[ H^{2}(\int_0^1w(s)dB_{s}^{2},1)\right] <\infty .$\newline
The two conditions above, together with
equation(\ref{theta_cond_2}), rule out doubling strategies.

\item $y\rightarrow H(y,t)$ is increasing for each fixed $t$, that is the
price increases if the stock demand increases.

\item $\mathbb{E}\left[ \left( h_i^{-1}\left( \mathbb{E}\left[ f(Z_{1})|\mathcal{F}^Z_{t_i}\right] \right)
\right) ^{2}\right] <\infty $ where $h_i^{-1}$ is the inverse of
$H(y,t_i)$.
\end{enumerate}

Moreover, $H$ is a rational pricing rule if, for a given $\theta $, it
satisfies
\begin{equation*}
H\left(\int_0^tw(s)dY_{s},t\right)=\mathbb{E}\left[
f(Z_{1})|\mathcal{F}_{t}^{M}\right] .
\end{equation*}
\end{definition}

\begin{remark}
Due to condition \textit{4} on the admissible pricing rules, the insider
can infer the total order process from the price process by inverting $%
H(\int_0^tw(s)dY_{s},t)=P_{t}$. Therefore, since I will be considering rational
 expectations equilibria, and because $w(s)$ is strictly positive, she can infer the total order process $Y_t$
and, since she knows her own total order process $\theta _{t}$, she
can deduce $B_{t}^{2}=Y_{t}-\theta _{t}$ from it. As a
consequence, the filtration of the insider can be written as $\mathcal{F}%
_{t}^{I}=\mathcal{F}_{t}^{B^{2},Z}=\mathcal{F}_{t}^{B^{2},B^{1}}\vee \sigma
(v)$, where $\sigma(v)$ is the sigma algebra generated by the random variable $v$.
\end{remark}

Given these definitions of admissible pricing rules and trading strategies,
it is now possible to formally define the market equilibrium as follows.

\begin{definition}
\label{eqm_def} A pair $((H^{\ast },w^{\ast}),\theta ^{\ast})$ is an
equilibrium if $ (H^{\ast },w^{\ast})$ is an admissible pricing
rule, $\theta ^{\ast }$ is admissible strategy, and:

\begin{enumerate}
\item Given $\theta ^{\ast }$, $(H^{\ast },w^{\ast})$ is a rational pricing rule, i.e.
it satisfies\newline
\begin{equation*}
H\left(\int_0^tw(s)dY_{s},t\right)=\mathbb{E}\left[
f(Z_{1})|\mathcal{F}_{t}^{M}\right] .
\end{equation*}

\item Given $(H^{\ast },w^{\ast})$, $\theta ^{\ast }$ solves the optimization problem%
\newline
\begin{equation*}
sup_{\theta \in \mathcal{A}(H^{\ast },w^{\ast})}\mathbb{E}\left[
(f(Z_{1})-P_{1})\theta _{1}+\int_{0}^{1}\theta _{s-}dP_{s}\right]
\end{equation*}%
\newline
Moreover, a pricing rule $(H^{\ast }(y,t),w^{\ast}(t))$ is an
inconspicuous equilibrium
pricing rule if there exists an inconspicuous insider trading strategy $%
\theta ^{\ast }$ such that $((H^{\ast },w^{\ast}),\theta ^{\ast })$
is an equilibrium.
\end{enumerate}
\end{definition}

Additionally, to define a well behaved problem I impose the following
technical conditions on the model parameters.

\begin{assumption}
\label{ass1} The fundamental value of the risky stock, $F(z,t)$, given by
\begin{equation}
F(Z_{t},t)=\mathbb{E}\left[ f(Z_{1})|\mathcal{F}_{t}^{Z}\right]
\label{fund_price}
\end{equation}%
is well defined and is a square integrable martingale, i.e.
\begin{equation}
\mathbb{E}\left[ f^{2}(Z_{1})\right] <\infty ,  \label{fin_val_cond}
\end{equation}%
and $f\left( .\right) $ is an increasing function.
\end{assumption}

\begin{assumption}
\label{ass2} The variance of the firm value, $\Sigma
_{z}(t)=\int_{0}^{t}\sigma _{z}^{2}(s)ds$, is finite for any $t$.
\end{assumption}

\begin{remark}
Since the final payoff of the stock is given by $f(Z_{1})$, the above
assumption implies that it is always possible to redefine the function $f$
so that%
\begin{equation}
\sigma ^{2}=1-\Sigma _{z}(1).
\end{equation}
In what follows, I will always assume that this equality holds.
\end{remark}

\section{The Markovian Equilibrium}
\setcounter{equation}{0}
In this section I address the problem of existence and uniqueness of
an equilibrium given by Definition \ref{eqm_def} in the case of
Markovian pricing rule i.e. I consider $w(t)\equiv
1$. Before stating the main result of this section, I need to impose
additional conditions on the model to insure that the problem is
well-posed.
\begin{assumption}
\label{ass3} For any $t\in \left[ 0,1\right) $ we have
\begin{equation}
\int_{0}^{t}\left( \Sigma _{z}(s)+\sigma ^{2}-s\right)
^{-2}ds<\infty \label{speed_of_adj}
\end{equation}%
and either
\begin{equation}
\int_{0}^{1}\left( \Sigma _{z}(s)+\sigma ^{2}-s\right)
^{-2}ds<\infty \label{speed_of_adj_1}
\end{equation}
or
\begin{equation}
\lim_{t\rightarrow 1}\int_{0}^{t}\frac{1}{\left\vert {\Sigma
_{z}(s)+\sigma ^{2}-s}\right\vert }ds=\infty
\label{speed_of_adj_infty}
\end{equation}
\end{assumption}

The above assumption is needed for the filtering problem of the
market maker to be well defined.

\begin{assumption}
\label{ass4}  There exists a $t^{\ast }\in \left[ 0,1\right) $ such
that
\begin{equation}\label{eq:signal_precision}
1-t>\int_{t}^{1}\sigma _{z}^{2}(s)ds
\end{equation} for any $t\geq t^{\ast }$ and $\sigma_z(t)$ is continuous on $[t^*,1]$.

Moreover, for all $t\in \left[ 0,1\right] $ we have
\begin{equation}
\Sigma _{z}(t)-t+\sigma ^{2}\geq 0.
\end{equation}
\end{assumption}
This assumption insures that: a) close to the market terminal time,
the insider's signal is more precise than the market maker's (i.e. $\mathbb{E}\left[ \left( Z_{1}-\mathbb{%
E}\left[ Z_{1}|\mathcal{F}_{t}^{M}\right] \right) ^{2}|\mathcal{F}_{t}^{M}%
\right] >\mathbb{E}\left[ \left( Z_{1}-\mathbb{E}\left[ Z_{1}|\mathcal{F}%
_{t}^{I}\right] \right) ^{2}|\mathcal{F}_{t}^{I}\right] $), and b)
that the insider's signal is always at least as precise as the
market maker's.
\begin{remark}
\label{speed_of_conv} Notice that Assumptions \ref{ass2}, \ref{ass3}
and \ref{ass4} guarantee that when condition (\ref{speed_of_adj_1})
is not satisfied
\begin{equation*}
\lambda (t)=\exp \left\{ -\int_{0}^{t}\frac{1}{\Sigma _{z}(s)+\sigma ^{2}-s}%
ds\right\} \overset{t\rightarrow 1}{\longrightarrow }0,
\end{equation*}%
and that if $\Xi (t)=\int_{0}^{t}\frac{1+\sigma _{z}^{2}(s)}{\lambda ^{2}(s)}%
ds\overset{t\rightarrow 1}{\longrightarrow }\infty $, then
\begin{equation}
\lim_{t\rightarrow 1}\lambda ^{2}(t)\Xi (t)\log \log \left( \Xi
(t)\right) =0.  \label{xi_cond}
\end{equation}%
Furthermore, assumption \ref{ass4} can be relaxed by replacing (\ref{eq:signal_precision}) with condition (\ref%
{xi_cond}).
\end{remark}

{\proof See Appendix A}

Now we are in the position to state the main result of this section
which is summarized in the next theorem.

{\theorem\label{eqm} Suppose that Assumptions \ref{ass1},
\ref{ass2}, \ref{ass3} and \ref{ass4} are satisfied. Then the pair
$(H^{\ast },\theta ^{\ast })$, where $H^{\ast }$ satisfies
\begin{eqnarray}
H_{t}(y,t)+\frac{1}{2}H_{yy}(y,t) &=&0  \label{price_HJB} \\
H(y,1) &=&f(y) , \label{price_HJB_fin}
\end{eqnarray}%
i.e. $H^{\ast}(y,t) = \mathbb{E}\left[f(y+B^2_1-B^2_t)\right]$ and
\begin{equation}
\theta _{t}^{\ast }=\int_{0}^{t}\frac{Z_{s}-Y_{s}}{\Sigma _{z}(s)-s+\sigma
^{2}}ds,  \label{opt_str}
\end{equation}%
is an equilibrium. Moreover, the pricing rule $H^{\ast }$ is the unique
inconspicuous equilibrium pricing rule in $\mathcal{H}$. Furthermore, given
this pricing rule $H^{\ast }$, the trading strategy $\theta^{\ast} $ is optimal in $%
\mathcal{A}(H^{\ast })$ for the insider if and only if }

\begin{enumerate}
\item \textit{The process $\theta^{\ast}_t$ is continuous and has bounded
variation.}

\item \textit{The total order, $Y_{t}^{\ast }=\theta _{t}^{\ast }+B_{t}^{2}$,
satisfies $Y_{1}^{\ast }=Z_{1}.$}
\end{enumerate}
Therefore, when the parameters of the market satisfy the stated
assumptions, there exists a unique Markovian pricing rule such that:
\emph{a)} at least one optimal trading strategy of the insider,
given by (\ref{opt_str}), is \emph{increasing} market efficiency
during \emph{all} trading periods since the insider pushes the price
to the fundamental value of the stock, \emph{b)} due to \emph{1},
the variance of the risky asset is not influenced by insider's
trading if she trades optimally, and \emph{c)} the insider presence
increases market efficiency close to the market termination time due
to \emph{2}.

I will prove this theorem in three propositions that focus on different
aspects of the equilibrium. In particular, the propositions will address:
\textit{a)} characterization of the optimal insider trading strategy, \textit{b)}
existence of the equilibrium, and \textit{c)} uniqueness of the inconspicuous
pricing rule.

The conclusion of Theorem \ref{eqm} is driven by the following
result: for any pricing rule in $\mathcal{H}$ satisfying equation
(\ref{price_HJB}), there exists a finite upper bound on the informed
agent's value function which is attained by a trading strategy which
is not detectable by the market maker, not locally correlated with
noisy trades, and such that all the private information is revealed
only at time $t=1$. Thus, this result gives the characterization of
the optimal insider's trading strategy in a slightly more general
form than stated in Theorem \ref{eqm}. This is summarized in the
following proposition.

{\proposition\label{char_tr_str} Suppose that Assumptions
\ref{ass1}, \ref{ass2}, \ref{ass3} and \ref{ass4} are satisfied.
Then, given an admissible pricing rule $H\in \mathcal{H} $
satisfying the partial differential equation (PDE)
(\ref{price_HJB}), an admissible trading strategy $\theta ^{\ast
}\in \mathcal{A}(H)$ is optimal for the insider if and only if: }

\begin{enumerate}
\item \textit{The process $\theta^{\ast}_t$ is continuous and has bounded
variation.}

\item \textit{The total order, $Y_{t}^{\ast }=\theta _{t}^{\ast }+B_{t}^{2}$,
satisfies}
\begin{equation}
h\left( Y_{1}^{\ast }\right) =H(Y_{1}^{\ast },1)=f\left(
Z_{1}\right),\label{final_cond}
\end{equation}
where $h(y)=H(y,1)$ and $f(Z_1)$ is the final payoff of the asset.
\end{enumerate}

\begin{proof}
\begin{itemize}
\item[(Sufficiency)] For any admissible trading strategy, by using
integration by parts for semimartingales (\cite{Pr}, Corollary II.6.2,
p. 68), we have
\begin{equation*}
\mathbb{E}\left[ X_{1}^{\theta }\right] =\mathbb{E}\left[
\int_{0}^{1}(F(Z_{s},s)-H(Y_{s-},s))d\theta _{s}+\int_{0}^{1}\theta
_{s-}dF(Z_{s},s)+\left[ \theta ,F(Z,\cdot )-H(Y,\cdot )\right] _{1}\right] .
\end{equation*}%
By applying It\^{o} formula for semimartingales (\cite{Pr}, Theorem
II.6.33, p. 81) to $H(y,t)$ and $F(z,t)$, and using the fact that $F(Z_{t},t)
$ is a true martingale, we get%
\begin{eqnarray*}
F(Z_{t},t) &=&F(z_{0},0)+\int_{0}^{t}F_{z}(Z_{s},s)dZ_{s} \\
H(Y_{t},t)
&=&H(0,0)+\int_{0}^{t}H_{y}(Y_{s-},s)dY_{s}+\int_{0}^{t}H_{t}(Y_{s}-,s)ds \\
&+&\frac{1}{2}\int_{0}^{t}H_{yy}(Y_{s-},s)d\left[ Y\right] _{s}+\sum_{s\leq
t}\left[ \Delta H(Y_{s},s)-H_{y}(Y_{s-},s)\Delta Y_{s}\right] .
\end{eqnarray*}%
Since $Y_{t}=\theta _{t}+B_{t}^{2}$, we have that $\left[ Y\right]
_{t}=t+\left\langle \theta ^{c}\right\rangle _{t}+2\left\langle \theta
^{c},B^{2}\right\rangle _{t}+\sum_{s\leq t}\left( \Delta \theta _{s}\right)
^{2}$. Therefore, using the fact that $H(y,t)$ satisfies equation (\ref%
{price_HJB}), we have that%
\begin{eqnarray*}
H(Y_{t},t) &=&H(0,0)+\int_{0}^{t}H_{y}(Y_{s-},s)dY_{s}^{c}+\frac{1}{2}%
\int_{0}^{t}H_{yy}(Y_{s-},s)d\left\langle \theta ^{c}\right\rangle _{s} \\
&+&\int_{0}^{t}H_{yy}(Y_{s-},s)d\left\langle \theta ^{c},B^{2}\right\rangle
_{s}+\sum_{s\leq t}\Delta H(Y_{s},s).
\end{eqnarray*}%
Therefore, by Theorem 26.6 of \cite{Kal}, and Theorem II.6.29 of \cite{Pr}, we have (notice that $Z_{s}$ and $B_{t}^{2}$ are continuous)%
\begin{eqnarray*}
\left[ \theta ,F(Z,\cdot )\right] _{1} &=&\int_{0}^{1}F_{z}(Z_{s},s)d\left[
\theta ^{c},Z\right] _{s} \\
\left[ \theta ,H(Y,\cdot )\right] _{1} &=&\int_{0}^{1}H_{y}(Y_{s-},s)d\left[
\theta ^{c}\right] _{s}+\int_{0}^{1}H_{y}(Y_{s-},s)d\left[ \theta ^{c},B^{2}%
\right] _{s}+\sum_{s\leq 1}\Delta H(Y_{s},s)\Delta \theta _{s}.
\end{eqnarray*}

On the other hand, consider a function
\begin{equation*}
J(y,z)=\int_{y}^{y^{\ast }(z)}\left( f(z)-H(x,1)\right) dx,
\end{equation*}%
where $y^{\ast }(z)$ is the solution of $H(y^{\ast }(z),1)=f(z)$. Let
\begin{equation}
V(y,z,t)=\mathbb{E}\left[ J\left( y+B_{1}^{2}-B_{t}^{2},z+\int_{t}^{1}\sigma
_{z}(s)dB_{s}^{1}\right) \right] .
\end{equation}%
This function is well defined (it is easy to check that $\mathbb{E}\left[
|J(B_{1}^{2},Z_{1})|\right] <\infty $) and satisfies the partial
differential equation
\begin{equation}
V_{t}(y,z,t)+\frac{1}{2}V_{yy}(y,z,t)+\frac{\sigma _{z}^{2}(t)}{2}%
V_{zz}(y,z,t)=0  \label{HJB1}
\end{equation}%
with terminal condition $V(y,z,1)=J(y,z)$. Therefore $V(y,z,1)\geq
V(y^{\ast }(z),z,1)=0$ for any fixed $z$ and any $y\neq y^{\ast
}(z)$. Moreover, since $H(y,1)$ is a nondecreasing continuous
function of $y$, we can use the monotone convergence theorem to
obtain
\begin{eqnarray*}
\lim_{\Delta \rightarrow 0+}\frac{V(y+\Delta ,z,t)-V(y,z,t)}{\Delta }
&=&\lim_{\Delta \rightarrow 0+}\frac{\mathbb{E}\left[ \int_{y+\Delta
+B_{1}^{2}-B_{t}^{2}}^{y+B_{1}^{2}-B_{t}^{2}}\left( f(z+\int_{t}^{1}\sigma
_{z}(s)dB_{s}^{1})-H(x,1)\right) dx\right] }{\Delta } \\
&=&-F(z,t)-\mathbb{E}\left[ \lim_{\Delta \rightarrow 0+}\frac{\int_{y+\Delta
+B_{1}^{2}-B_{t}^{2}}^{y+B_{1}^{2}-B_{t}^{2}}H(x,1)dx}{\Delta }\right]  \\
&=&\mathbb{E}\left[ H(y+B_{1}^{2}-B_{t}^{2},1)\right] -F(z,t).
\end{eqnarray*}%
Thus, due to the definition of an admissible pricing rule, we have
\begin{equation}
\lim_{\Delta \rightarrow 0+}\frac{V(y+\Delta ,z,t)-V(y,z,t)}{\Delta }%
+F(z,t)-H(y,t)=0.
\end{equation}%
The same argument can be applied to the left derivative of $V$ with respect
to $y$ to obtain
\begin{equation}
V_{y}(y,z,t)+F(z,t)-H(y,t)=0.  \label{HJB2}
\end{equation}%
As a consequence, we can express $\mathbb{E}\left[ X_{1}^{\theta }\right] $
in terms of $V$ as (notice that $\int_{0}^{t}B_{t}^{2}dF(Z_{t},t)$ is a
martingale)
\begin{eqnarray*}
\mathbb{E}\left[ X_{1}^{\theta }\right]  &=&\mathbb{E}\left[
-\int_{0}^{1}V_{y}(Y_{s-},Z_{s},s)d\theta
_{s}-\int_{0}^{1}V_{z}(Y_{s-},Z_{s},s)dZ_{s}-%
\int_{0}^{1}V_{zy}(Y_{s-},Z_{s},s)d\left[ \theta ^{c},Z\right] _{s}\right.
\\
&-&\left. \int_{0}^{1}V_{yy}(Y_{s-},Z_{s},s)d\left[ \theta ^{c}\right]
_{s}-\int_{0}^{1}V_{yy}(Y_{s-},Z_{s},s)d\left[ \theta ^{c},B^{2}\right]
_{s}-\sum_{s\leq 1}\Delta V_{y}(Y_{s},Z_{s},s)\Delta \theta _{s}\right] .
\end{eqnarray*}%
On the other hand, by applying the It\^{o} formula for semimartingales to $V$
directly (\cite{Pr}, Theorem II.6.33, p. 81) we get
\begin{eqnarray*}
\mathbb{E}\left[ V(Y_{1},Z_{1},1)\right]  &=&\mathbb{E}\left[
V(0,Z_{0},0)-X_{1}^{\theta
}+\int_{0}^{1}V_{y}(Y_{s-},Z_{s},s)dB_{s}^{2}\right.  \\
&-&\left. \frac{1}{2}\int_{0}^{1}V_{yy}(Y_{s-},Z_{s},s)d\left[ \theta ^{c}%
\right] _{s}+\sum_{s\leq 1}\left[ \Delta
V(Y_{s},Z_{s},s)-V_{y}(Y_{s},Z_{s},s)\Delta Y_{s}\right] \right] .
\end{eqnarray*}%
Notice that, due to the definition of the fundamental value $F$ and of
admissible pricing rule $H$, we have $\mathbb{E}\left[
\int_{0}^{1}V_{z}(Y_{s},Z_{s},s)dB_{s}^{2}\right] =0$. Therefore
\begin{eqnarray*}
\mathbb{E}\left[ X_{1}^{\theta }\right]  &=&\mathbb{E}\left[
V(0,Z_{0},0)-V(Y_{1},Z_{1},1)-\frac{1}{2}\int_{0}^{1}V_{yy}(Y_{s-},Z_{s},s)d%
\left[ \theta ^{c}\right] _{s}\right.  \\
&+&\left. \sum_{s\leq 1}\left[ \Delta
V(Y_{s},Z_{s},s)-V_{y}(Y_{s},Z_{s},s)\Delta Y_{s}\right] \right] .
\end{eqnarray*}%
Moreover, due to the properties of $V$ we have
\begin{eqnarray}
\sum_{s\leq 1}\left( \Delta V(Y_{s},Z_{s},s)-V_{y}(Y_{s},Z_{s},s)\Delta
Y_{s}\right)  &\leq &0,  \label{jump} \\
-\int_{0}^{1}\frac{V_{yy}(Y_{s},Z_{s},s)}{2}d\left[ \theta ^{c}\right] _{s}
&\leq &0 , \label{mart_part} \\
-V(1,Y_{1},Z_{1}) &\leq &-V(1,y^{\ast }(Z_{1}),Z_{1}).  \label{final}
\end{eqnarray}%
The above inequalities become equalities if and only if the following
conditions hold: $\Delta \theta =0$ for equation  (\ref{jump}); $\left[
\theta ^{c}\right] _{1}=0$ for equation (\ref{mart_part}); $H(Y_{1}^{\ast
},1)=f\left( Z_{1}\right) $ for equation (\ref{final}).

Therefore, for any function $V$ satisfying equations (\ref{HJB1}), (\ref%
{HJB2}) and the final condition given by $V(y,z,1)\geq V(y^{\ast }(z),z,1)=0$
for every $z$ and any $y\neq y^{\ast }(z)$ (where $y^{\ast }(z)$ is the
solution of $H(y^{\ast }(z),1)=f(z)$), we have that
\begin{equation*}
\mathbb{E}\left[ X_{1}^{\theta }\right] \leq V(0,Z_{0},0).
\end{equation*}%
This expression holds with equality if and only if $\theta $ is continuous
and condition (\ref{final_cond}) is satisfied. Hence, if $\theta $ is such
that these conditions are satisfied, then it is optimal.

\item[(Necessity)] Consider the continuous martingale given by
\begin{equation*}
X_{t}=G(Z_{t},t)=\mathbb{E}\left[ h^{-1}(f(Z_{1}))|\mathcal{F}_{t}^{I}\right]
.
\end{equation*}%
This martingale is well defined since $H$ is an admissible pricing rule.%
\newline
Consider $\theta _{t}=\int_{0}^{t}\frac{X_{s}-Y_{s}}{1-s}ds$. In this case,
we can solve the stochastic differential equation for $Y$ to get
\begin{equation*}
Y_{t}=X_{t}-\left( 1-t\right) \left( v+\int_{0}^{t}\frac{1}{1-s}%
dX_{s}-\int_{0}^{t}\frac{1}{1-s}dB_{s}^{2}\right) .
\end{equation*}%
Notice that $Y_{t}$ is continuous, therefore $\theta _{t}$ has bounded
variation almost surely. Moreover, $H(Y_{1},1)=f(Z_{1})$ almost surely,
hence this choice of $\theta $ gives
\begin{equation*}
\mathbb{E}\left[ X_{1}^{\theta }\right] =V(0,Z_{0},0).
\end{equation*}%
Since, by the sufficiency proof, we have that for any
$\tilde{\theta}_{t}$ which is either not continuous or does not
satisfy equation (\ref{final_cond})
\begin{equation*}
\mathbb{E}\left[ X_{1}^{\tilde{\theta}}\right] <V(0,Z_{0},0)=\mathbb{E}\left[
X_{1}^{\theta }\right] ,
\end{equation*}%
we know that any such $\tilde{\theta}_{t}$ is not optimal.
\end{itemize}
\end{proof}

From this characterization result, it follows that the $\theta _{t}^{\ast }$
given by (\ref{opt_str}) is an optimal insider trading strategy given an
admissible pricing rule $H^{\ast }$ satisfying (\ref{price_HJB}) and (\ref%
{price_HJB_fin}). Establishing the rationality of the pricing rule $H^{\ast }
$, on the other hand, is not so direct. Therefore, to set up the stage for
proving that the $(H^{\ast },\theta ^{\ast })$ given in Theorem \ref{eqm} is
indeed an equilibrium, we first need to demonstrate the following lemma.

{\lemma\label{filt} Consider the process $Y_{t}$ satisfying the stochastic
differential equation
\begin{equation*}
dY_{s}=\frac{Z_{s}-Y_{s}}{\Sigma _{z}(s)-s+\sigma ^{2}}ds+dB_{s}^{2},
\end{equation*}%
with
\begin{equation*}
Z_{t}=v+\int_{0}^{t}\sigma _{z}(s)dB_{s}^{1},
\end{equation*}%
where $B_{t}^{1}$ and $B_{t}^{2}$ are two independent standard Brownian
motions, $v$ is $N(0,\sigma )$ independent of $\mathcal{F}_{1}^{B^{1},B^{2}}$
and $\Sigma _{z}(t)=\int_{0}^{t}\sigma _{z}^{2}(s)ds$. Suppose that $\sigma $%
, $\sigma _{z}(t)$ and $\Sigma _{z}(t)$ satisfy Assumptions
\ref{ass1}, \ref{ass2}, \ref{ass3} and \ref{ass4}. Then, on the
filtration $\bar{\mathcal{F}}_{t}^{Y}$, the process $Y_{t}$ is a
standard Brownian motion and $Y_{1}=Z_{1}$.}

\begin{proof}
Fix any $T\in \lbrack 0,1)$. From Theorem 10.3 of \cite{L} (note that, due to Assumption {\ref{ass3}}, the
conditions of the
theorem are satisfied), we have that on the filtration $\left( \mathcal{F}%
_{t}^{Y}\right) _{t\leq T}$ the stochastic differential equation for $Y$ is
\begin{equation*}
dY_{s}=\frac{m_{s}-Y_{s}}{\Sigma _{z}(s)-s+\sigma ^{2}}ds+dB_{s}^{Y},
\end{equation*}%
with
\begin{equation*}
dm_{s}=\frac{\gamma _{s}}{\Sigma _{z}(s)-s+\sigma ^{2}}dB_{s}^{Y},
\end{equation*}%
where $B_{t}^{Y}$ is Brownian motion on $\mathcal{F}_{t}^{Y}$, and $\gamma
_{s}$ satisfies the following ordinary differential equation (ODE)
\begin{equation*}
\dot{\gamma _{s}}=\sigma _{z}^{2}(s)-\frac{\gamma _{s}^{2}}{\left(
\Sigma _{z}(s)-s+\sigma ^{2}\right) ^{2}}
\end{equation*}%
with initial condition $\gamma _{0}=\sigma ^{2}$.

Notice that $\gamma _{s}=\Sigma _{z}(s)-s+\sigma ^{2}$ is the unique
solution of this ODE and initial condition. Therefore on $\left( \mathcal{F}%
_{t}^{Y}\right) _{t\leq T}$, the process $Y$ satisfies
\begin{equation*}
dY_{s}=\frac{B_{s}^{Y}-Y_{s}}{\Sigma _{z}(s)-s+\sigma ^{2}}ds+dB_{s}^{Y}.
\end{equation*}%
The unique strong solution of this stochastic differential equation on $[0,T]
$ is $Y_{s}=B_{s}^{Y}$ (see \cite{Ka2}, Example 5.2.4).
Hence, on the interval $[0,1)$, the process $Y$ is a Brownian motion on its
own (completed) filtration. By continuity of $Y$, this process is a Brownian
motion on $[0,1]$. To prove that $Y_{1}=Z_{1}$, notice that
\begin{equation*}
Y_{t}^{\ast }=Z_{t}+\lambda (t)\left( -v+\int_{0}^{t}\frac{1}{\lambda (s)}%
dB_{s}^{2}-\int_{0}^{t}\frac{\sigma _{z}(s)}{\lambda
(s)}dB_{s}^{1}\right)
\end{equation*}%
where $\lambda (t)=\exp \left\{ -\int_{0}^{t}\frac{1}{\Sigma _{z}(s)+\sigma
^{2}-s}ds\right\} $.

Note that a random variable $\int_{0}^{t}\frac{1}{\lambda (s)}%
dB_{s}^{2}-\int_{0}^{t}\frac{\sigma _{z}(s)}{\lambda (s)}dB_{s}^{1}$
is normally distributed with mean $0$ and variance
$\int_{0}^{t}\frac{1+\sigma
_{z}^{2}(s)}{\lambda ^{2}(s)}ds$. Therefore, due to the Assumption {\ref%
{ass3}} (and in particular condition (\ref{speed_of_adj_infty})), if $%
\lim_{t\rightarrow 1}\int_{0}^{t}\frac{1+\sigma _{z}^{2}(s)}{\lambda ^{2}(s)}%
ds<\infty $, then $Y_{1}=Z_{1}$.

On the other hand, if $\lim_{t\rightarrow 1}\int_{0}^{t}\frac{1+\sigma
_{z}^{2}(s)}{\lambda ^{2}(s)}ds=\infty $, consider the process
\begin{equation*}
X_{t}=\int_{0}^{t}\frac{1}{\lambda
(s)}dB_{s}^{2}-\int_{0}^{t}\frac{\sigma _{z}(s)}{\lambda
(s)}dB_{s}^{1},
\end{equation*}%
and a change of time $\tau (t)$ given by
\begin{equation*}
\int_{0}^{\tau (t)}\frac{1+\sigma _{z}^{2}(s)}{\lambda ^{2}(s)}ds=t.
\end{equation*}%
Then, $W_{s}=X_{\tau (s)}$ is a Brownian motion. Hence, we can use the law
of iterated logarithm to get
\begin{eqnarray*}
\limsup_{s\rightarrow \infty }\frac{W_{s}}{\sqrt{2s\log \log s}} &=&1 \\
\liminf_{s\rightarrow \infty }\frac{W_{s}}{\sqrt{2s\log \log s}} &=&-1
\end{eqnarray*}%
or, in the original time,
\begin{eqnarray*}
\limsup_{t\rightarrow 1 }\frac{X_{t}}{\sqrt{2\Xi (t)\log \log (\Xi (t)})%
} &=&1 \\
\liminf_{t\rightarrow 1 }\frac{X_{t}}{\sqrt{2\Xi (t)\log \log (\Xi (t)})%
} &=&-1
\end{eqnarray*}%
where $\Xi (t)=\int_{0}^{t}\frac{1+\sigma _{z}^{2}(s)}{\lambda
^{2}(s)}ds$. Since, due to the Assumptions \ref{ass2},
\ref{ass3} and \ref{ass4}, we have
\begin{equation*}
\lim_{t\rightarrow 1}\lambda ^{2}(t)\Xi (t)\log \log \left( \Xi (t)\right)
=0,
\end{equation*}%
it follows that $\lim_{t\rightarrow 1}\lambda (t)X_{t}=0$, therefore $%
Y_{1}=Z_{1}$.
\end{proof}

With this lemma at hand, establishing that the pair $(H^{\ast },\theta
^{\ast })$ given in the Theorem \ref{eqm} is indeed an equilibrium is
straightforward, as the following proposition demonstrates.

{\proposition\label{equilibrium} Suppose that Assumptions
\ref{ass1}, \ref{ass2}, \ref{ass3} and \ref{ass4} are satisfied.
Then the pair $(H^{\ast },\theta ^{\ast })$, where
$H^{\ast }(y,t)$ satisfies the partial differential equation (PDE) (\ref%
{price_HJB}) with terminal condition (\ref{price_HJB_fin}), and the process $%
\theta _{t}^{\ast }$ is given by (\ref{opt_str}), is an equilibrium.}

\begin{proof}
Due to Proposition \ref{char_tr_str}, the $\theta _{t}^{\ast }$ defined by (%
\ref{opt_str}) is the optimal trading strategy given the admissible pricing
rule $H^{\ast }(y,t)$ which satisfies equation (\ref{price_HJB}) and (\ref%
{price_HJB_fin}), if and only if: \textit{a)} $\theta _{t}^{\ast }$ is
continuous with bounded variation, and \textit{b)} $Y_{t}^{\ast }=\theta
_{t}^{\ast }+B_{t}^{2}$ satisfies $Y_{1}^{\ast }=Z_{1}$. Due to Lemma \ref%
{filt}, we have that $\theta _{t}^{\ast }$ is continuous with bounded
variation, and $Y_{1}^{\ast }=Z_{1}$. Therefore $\theta _{t}^{\ast }$ is an
optimal trading strategy given the pricing rule $H^{\ast }(y,t)$.

On the other hand, due to the Lemma \ref{filt}, for $\theta ^{\ast }$ given
by (\ref{opt_str}), $Y^{\ast }$ is a Brownian motion with $Y_{1}^{\ast
}=Z_{1}$. Therefore, the rational pricing rule given $\theta ^{\ast }$
should be
\begin{equation*}
H(y,t)=\mathbb{E}\left[ f(y+B_{1}^{2}-B_{t}^{2})\right] .
\end{equation*}%
This pricing rule satisfies the PDE (\ref{price_HJB}) with terminal
condition (\ref{price_HJB_fin}). Therefore, $H^{\ast }(y,t)=H(y,t)$
is a rational pricing rule. Hence, the pair $(H^{\ast },\theta
^{\ast })$ given in this proposition is an equilibrium.
\end{proof}

To complete the proof of Theorem \ref{eqm}, we need to show uniqueness
of the inconspicuous pricing rule in $\mathcal{H}$.

{\proposition \label{equilibrium} The pricing rule $H^{\ast}(y,t)$ which
satisfies the PDE (\ref{price_HJB}), with terminal condition (\ref%
{price_HJB_fin}), is the unique inconspicuous pricing rule.}

\begin{proof}
From Proposition \ref{equilibrium} and Lemma \ref{filt}, it directly
follows that $H^{\ast }(y,t)$ satisfying the PDE (\ref{price_HJB})
with terminal condition (\ref{price_HJB_fin}) is an inconspicuous
equilibrium pricing rule. To prove uniqueness, consider some
equilibrium inconspicuous pricing rule $H$. By definition, there
exists a trading strategy $\theta _{t}\in \mathcal{A}(H)$ such that
the  $(H,\theta )$ is an equilibrium, and the
total order process $Y_{t}=\theta _{t}+B_{t}^{2}$ is a Brownian motion on $%
\mathcal{F}_{t}^{M}$. By the definition of equilibrium,
\begin{equation*}
H(Y_{t},t)=\mathbb{E}\left[ f(Z_{1})|\mathcal{F}_{t}^{M}\right] =\mathbb{E}%
\left[ H(Y_{1},1)|\mathcal{F}_{t}^{M}\right] .
\end{equation*}%
Since $Y_{t}$ is a Brownian motion on $\mathcal{F}_{t}^{M}$, and given the
definition of admissible pricing rule, $H$ must satisfy the PDE (\ref%
{price_HJB}) with terminal condition $H(y,1)=h(y)$, for some nondecreasing
function $h$ with $\mathbb{E}\left[ h^{2}(Y_{1})\right] <\infty $. Hence, to
show uniqueness of $H^{\ast }$ we need to demonstrate that $h=f$ almost
everywhere. Due to Proposition \ref{char_tr_str}, it follows from the
optimality of $\theta $ that $f(Z_{1})=h(Y_{1})$ and, since $\theta $ is
inconspicuous, $Y_{1}\sim N(0,1)$. Since $Z_{1}\sim N(0,1)$ by definition,
one can have $f(Z_{1})=h(Y_{1})$ if and only if $f=h$ almost everywhere,
hence $H^{\ast }$ is indeed a unique inconspicuous pricing rule.
\end{proof}
\section{Non Markovian equilibrium}
\setcounter{equation}{0}
In this section I address the problem of existence of an equilibrium
given by Definition \ref{eqm_def} in the more general case of non
Markovian pricing rule, i.e. I consider general weighting functions
$w(t)$ satisfying Definition \ref{pr_rule_def} thus allowing the market
maker to assign different weights to the information she receives.

As in the case of Markovian pricing rule, the existence of an
equilibrium result  is driven by the existence of  a finite upper bound on
the informed agent's value function, and the characterization of the
trading strategies which attain it. This characterization is
summarized in the following proposition.

{\proposition\label{char_tr_str_nm} Suppose that Assumptions
\ref{ass1} and \ref{ass2} are satisfied. Then, given an admissible
pricing rule $w(t)$ defined by (\ref{def:weight_func}) with
$\sigma_y^i<\sigma_y^{i+1}$ for any $i$ and $(H,w)\in \mathcal{H} $
with $H$ satisfying the partial differential equation
\begin{equation}\label{price_HJB_nm}
H_{t}(y,t)+\frac{w^2(t)}{2}H_{yy}(y,t) =0
\end{equation}
an admissible trading strategy $\theta ^{\ast }\in \mathcal{A}(H,w)$
is optimal for insider if and only if: }
\begin{enumerate}
\item \textit{The process $\theta^{\ast}_t$ is continuous and has bounded
variation.}
\item \textit{The weighted total order, $\xi_{t}^{\ast }=\int_0^tw(s)d\theta _{s-}^{\ast }+\int_0^tw(s)dB_{s}^{2}$
satisfies}
\begin{equation}
h_i\left( \xi_{t_i}^{\ast }\right)= H(\xi_{t_i}^{\ast },t_i)=F\left(
Z_{t_i},t_i\right) . \label{final_cond_nm}
\end{equation}
\end{enumerate}
Therefore, as in the case of Markovian pricing rule, the optimal
strategy of the insider does not alter quadratic variation of total
order process, does not add jumps to it and is uncorrelated with it.
But, differently from the Markovian case, it follows from
(\ref{final_cond_nm}) that in this setting it is \emph{optimal} for
the insider to reveal her information not only at the market
terminal time, but also in the interim times whenever the market maker
changes her weighting function.

\begin{proof}
\begin{itemize}
\item[(Sufficiency)]   As in the proof of Proposition \ref{char_tr_str}, for any admissible trading strategy we have
\begin{eqnarray*}
\mathbb{E}\left[ X_{1}^{\theta }\right] = \mathbb{E}\left[
\int_{0}^{1}(F(Z_{s},s)-H(\xi_{s-},s))d\theta
_{s}+\int_{0}^{1}\theta
_{s-}dF(Z_{s},s)+\int_{0}^{1}F_{z}(Z_{s},s)d\left[ \theta
^{c},Z\right] _{s}\right.\\
\left.  -  \int_{0}^{1}H_{\xi}(\xi_{s-},s)w(s)d\left[
\theta ^{c}\right] _{s}-\int_{0}^{1}H_{\xi}(\xi_{s-},s)w(s)d\left[ \theta ^{c},B^{2}%
\right] _{s}-\sum_{s\leq 1}\Delta H(\xi_{s},s)\Delta \theta
_{s}\right].
\end{eqnarray*}

On the other hand, consider the functions
\begin{equation*}
J^i(\xi,z)=\int_{\xi}^{\xi^{\ast }(z)}\left(
F(z,t_i)-H(x,t_i)\right) dx,
\end{equation*}%
where $\xi_i^{\ast }(z)$ is the solution of $H(\xi_i^{\ast
}(z),t_i)=F(z,t_i)$. For $t\leq t_i$ let
\begin{equation*}
V^i(\xi,z,t)=\mathbb{E}\left[ J^i\left(
\xi+\int_{t}^{t_i}w(s)dB_{s}^{2},z+\int_{t}^{t_i}\sigma
_{z}(s)dB_{s}^{1}\right) \right].
\end{equation*}
These functions are well defined (it is easy to check that
$\mathbb{E}\left[ |J^i(\int_{0}^{t_i}w(s)dB_{s}^{2},Z_{t_i})|\right]
<\infty $) and satisfy the partial differential equation
\begin{equation*}
V^i_{t}(\xi,z,t)+\frac{w^2(s)}{2}V^i_{\xi\xi}(\xi,z,t)+\frac{\sigma
_{z}^{2}(t)}{2}V^i_{zz}(\xi,z,t)=0
\end{equation*}%
with terminal condition $V^i(\xi,z,t_i)=J^i(\xi,z)$. Therefore,
$V^i(\xi,z,t_i)\geq V(\xi_i^{\ast }(z),z,t_i)=0$ for any fixed $z$
and any $\xi\neq \xi_i^{\ast }(z)$. Moreover, since $H(\xi,t)$ is a
nondecreasing continuous function of $\xi$, and due to the definition
of an admissible pricing rule, we have
\begin{equation*}
V^i_{\xi}(\xi,z,t)+F(z,t)-H(\xi,t)=0.
\end{equation*}
Define the function $V$ as
\begin{equation*}
V(\xi,z,t)=\sum_{i<n:t\leq
t_i}\left(\frac{1}{\sigma_y^i}-\frac{1}{\sigma_y^{i+1}}\right)V^i(\xi,z,t)+\frac{1}{\sigma_y^n}V^n(\xi,z,t)
\end{equation*}
Notice that, due to the properties of the functions $V^i$, we have that
$V$ is well defined and satisfies the partial differential equation
\begin{equation}
V_{t}(\xi,z,t)+\frac{w^2(s)}{2}V_{\xi\xi}(\xi,z,t)+\frac{\sigma
_{z}^{2}(t)}{2}V_{zz}(\xi,z,t)=0,  \label{HJB1_nm}
\end{equation}%
with conditions
$V(\xi,z,t_i)=\left(\frac{1}{\sigma_y^i}-\frac{1}{\sigma_y^{i+1}}\right)J^i(\xi,z)+V(\xi,z,t_i+)$
if $i<n$ and $V(\xi,z,t_n)=\frac{1}{\sigma_y^n}J^n(\xi,z)$.
Moreover, we have
\begin{equation}
V_{\xi}(\xi,z,t)+\frac{F(z,t)-H(\xi,t)}{w(t)}=0. \label{HJB2_nm}
\end{equation}
As a consequence, we can express $\mathbb{E}\left[ X_{1}^{\theta
}\right] $ in terms of $V$ as (notice that $\int_{0}^{t}\int_0^u
w(s)dB_{s}^{2}dF(Z_{u},u)$ is a martingale)
\begin{eqnarray*}
\mathbb{E}\left[ X_{1}^{\theta }\right]  &=&\mathbb{E}\left[
-\int_{0}^{1}V_{\xi}(\xi_{s-},Z_{s},s)w(s)d\theta
_{s}-\int_{0}^{1}V_{z}(\xi_{s-},Z_{s},s)dZ_{s}\right.
\\
&-&\left. \int_{0}^{1}V_{z\xi}(\xi_{s-},Z_{s},s)w(s)d\left[ \theta
^{c},Z\right] _{s} -
\int_{0}^{1}V_{\xi\xi}(\xi_{s-},Z_{s},s)w^2(s)d\left[ \theta
^{c}\right] _{s}\right. \\
&-& \left.\int_{0}^{1}V_{\xi\xi}(\xi_{s-},Z_{s},s)w^2(s)d\left[
\theta ^{c},B^{2}\right] _{s}-\sum_{s\leq 1}\Delta
(w(s)V_{\xi}(\xi_{s},Z_{s},s))\Delta \theta _{s}\right] .
\end{eqnarray*}%
On the other hand, by applying the It\^{o} formula for
semimartingales to $V$ directly (\cite{Pr}, Theorem II.6.33, p.
81) and removing martingale terms we get
\begin{eqnarray*}
\mathbb{E}\left[ X_{1}^{\theta }\right]  &=&\mathbb{E}\left[
V(0,Z_{0},0)-\sum_{i=1}^{n-1}\left(\frac{1}{\sigma_y^i}-\frac{1}{\sigma_y^{i+1}}\right)J^i(\xi_{t_i},Z_{t_i},t_i)-\frac{1}{\sigma_y^n}J^n(\xi_{t_i},Z_{t_i},t_i)\right.\\
&-& \frac{1}{2}\int_{0}^{1}V_{\xi\xi}(\xi_{s-},Z_{s},s)w^2(s)d\left[
\theta ^{c}\right] _{s} +\left. \sum_{s\leq 1}\left[ \Delta
V(\xi_{s},Z_{s},s)-V_{\xi}(\xi_{s},Z_{s},s)\Delta \xi_{s}\right]
\right] .
\end{eqnarray*}%
Moreover, due to the properties of $V$ we have
\begin{eqnarray}
\sum_{s\leq 1}\left( \Delta
V(\xi_{s},Z_{s},s)-V_{\xi}(\xi_{s},Z_{s},s)\Delta
\xi_{s}\right)  &\leq &0,  \label{jump_nm} \\
-\int_{0}^{1}\frac{V_{\xi\xi}(\xi_{s},Z_{s},s)w^2(s)}{2}d\left[
\theta ^{c}\right] _{s}
&\leq &0,  \label{mart_part_nm} \\
-J^i(t_i,\xi_{t_i},Z_{t_i}) &\leq &0. \label{final_nm}
\end{eqnarray}%
The above inequalities become equalities if and only if the
following conditions hold: $\Delta \theta =0$ for equation
(\ref{jump_nm}); $\left[ \theta ^{c}\right] _{1}=0$ for equation
(\ref{mart_part_nm}); $H(\xi_{t_i}^{\ast },t_i)=F\left(
Z_{t_i},t_i\right) $ for equations (\ref{final_nm}).

Therefore, we have that
\begin{equation*}
\mathbb{E}\left[ X_{1}^{\theta }\right] \leq V(0,Z_{0},0).
\end{equation*}%
This expression holds with equality if and only if $\theta $ is
continuous with bounded variation and condition
(\ref{final_cond_nm}) is satisfied. Hence, if $\theta $ is such that
these conditions are satisfied, then it is optimal. 
\item[(Necessity)] Consider the process given by
\begin{equation*}
X_{t}=G(Z_{t},t)=\sum_{i=1}^n\mathbb{E}\left[
h_i^{-1}(F(Z_{t_i},t_i))|\mathcal{F}_{t}^{I}\right]1_{\{t\in(t_{i-1},t_i]\}}
\end{equation*}
with $X(0)=\mathbb{E}\left[
h_1^{-1}(F(Z_{t_1},t_1))|\mathcal{F}_{0}^{I}\right]$ where
$h^{-1}_i$ is inverse of $H(y,t_i)$.
This process is well defined since $H$ is an admissible pricing rule.%
\newline
Consider the  trading strategy given by  $\theta_0=0$ and $d\theta
_{t}=\sum_{i=1}^n\frac{X_{t}-\xi_{t}}{\sigma_y^i\left(t_{i}-s\right)}1_{\{t\in (t_{i-1},t_i]\}}dt$.
In this case, we can solve the stochastic differential equation for
$\xi$ on each interval $[t_{i-1},t_{i}]$ to get
\begin{equation*}
\xi_{t}=X_{t}-\left( t_i-t\right) \left( \frac{X_{t_{i-1}}-\xi_{t_{i-1}}}{t_i-t_{i-1}}+\int_{t_{i-1}}^{t}\frac{1}{t_i-s}
dX_{s}-\int_{t_{i-1}}^{t}\frac{\sigma_y^i}{t_i-s}dB_{s}^{2}\right) .
\end{equation*}%
Notice that $\xi_{t}$ is finite almost surely, therefore $\theta _{t}$ has
bounded variation almost surely. Moreover,
$H(\xi_{t_i},t_1)=F(Z_{t_i},t_i)$ almost surely, hence this choice
of $\theta $ gives
\begin{equation*}
\mathbb{E}\left[ X_{1}^{\theta }\right] =V(0,Z_{0},0).
\end{equation*}%
Since, by the sufficiency proof, we have that for any
$\tilde{\theta}_{t}$ which is either not continuous or does not
satisfy equation (\ref{final_cond})
\begin{equation*}
\mathbb{E}\left[ X_{1}^{\tilde{\theta}}\right]
<V(0,Z_{0},0)=\mathbb{E}\left[ X_{1}^{\theta }\right] ,
\end{equation*}%
we know that any such $\tilde{\theta}_{t}$ is not optimal.
\end{itemize}
\end{proof}

From this characterization follows the existence of equilibrium
result, as the next theorem demonstrates.

{\theorem\label{th:eqm_nm} Suppose that $\sigma_z(t)$ and $\sigma$
are such that there exists a piecewise constant function
$g(t)=\sum_{i=1}^{n}\alpha_i 1_{\{t\in(t_{i-1},t_i]\}}$ with
$0=t_0<\ldots<t_n=1$, $0<\alpha_i<\alpha_{i+1}$ for any $i$ and
$\sum_{i=1}^{n}\alpha^2_i=1$, satisfying the following conditions:
\begin{eqnarray}\label{eq:sigma_c1}
\Sigma_z(t)+\sigma^2-\int_0^t g^2(s)ds & > & 0 \mbox{ for all
$t\in [0,1]\backslash{\{t_i\}}_{i=0}^n$,}\\
\label{eq:sigma_c2}\Sigma_z(t_i)+\sigma^2-\int_0^{t_i} g^2(s)ds &=&
0 \mbox{ for all $t_i$,}\\
\label{eq:sigma_c3}
\int_{t_{i-1}}^{t}\frac{1}{\left(\Sigma_z(s)+\sigma^2-\int_0^s
g^2(u)du\right)^2}ds &<& \infty \mbox{ for all $t\in [t_{i-1},t_i)$
and any $i\leq n$,}\\\label{eq:sigma_c4} \lim_{t\rightarrow
t_i}\int_{t_{i-1}}^{t}\frac{1}{\Sigma_z(s)+\sigma^2-\int_0^s
g^2(u)du}ds &=& \infty.
\end{eqnarray}
Then there exists an equilibrium and it is given by the weighting
function $w^{\ast}(s)=g(s)$, the pricing rule
$H^{\ast}(\xi,t)=\mathbb{E}\left[f\left(\xi+\int_t^1g(s)dB^2_s\right)\right]$, and the trading
strategy $\theta^{\ast}_t$ satisfying $\theta^{\ast}_0=0$ and
\begin{equation*}
d\theta^{\ast}_t=1_{\{t\in(0,t_{1}]\}}\frac{\left(Z_t-\alpha_{1}Y_t\right)\alpha_{1}}{\Sigma_z(t)+\sigma^2-\int_0^{t}
g^2(s)ds }dt+\sum_{i=1}^{n-1}1_{\{t\in(t_i,t_{i+1}]\}}
\frac{\left(Z_t-Z_{t_i}-\alpha_{i+1}\left(Y_t-Y_{t_i}\right)\right)\alpha_{i+1}}{\Sigma_z(t)+\sigma^2-\int_0^{t}
g^2(s)ds }dt.
\end{equation*}
}

This theorem implies that if there are times $t_i$ such that
$\Sigma_z(t_i)+\sigma^2-\int_0^{t_i} w^2(s)ds = 0$, and the
intensity of private information arrival is fast enough at these
points (i.e. (\ref{eq:sigma_c3}) is satisfied), then it is:
\emph{a)} rational for the market maker to change her weighting
function at these points and \emph{b)} it is optimal for the insider
to reveal her information at these times.

Moreover, notice that this equilibrium exists even when assumptions
\ref{ass3} and \ref{ass4} insuring existence of Markovian equilibrium are not satisfied. That is, even if
$\Sigma_z(s)+\sigma^2-s<0$ for some $s$, there is a non Markovian
equilibrium as long as there exists a piecewise linear increasing
function $g$, the integral of which is bounding the realized
variance of the insider signal ($\Sigma_z(t)-\sigma^2$) from below and
satisfies the conditions of the theorem.

The proof of this theorem relies on linear filtering and
deterministic time change.

\begin{proof}
To demonstrate that $((H^{\ast},w^{\ast}),\theta^{\ast})$ is an
equilibrium, it is enough to show that
\begin{equation}\label{eq:cond1_nm}
Y^{\ast}_t-Y^{\ast}_{t_i} \mbox{ is a Brownian motion on
$[t_i,t_{i+1}]$ in its own filtration},
\end{equation}
where $Y^{\ast}_t=\theta^{\ast}_t+B^2_t$  and
\begin{equation}\label{eq:cond2_nm}
\alpha_{i+1} \left(Y_{t_{i+1}}-Y_{t_i}\right)=Z_{t_{i+1}}-Z_{t_{i}}.
\end{equation}
Indeed, if these two conditions are satisfied, since
$\Sigma_z(t_i)+\sigma^2-\int_0^{t_i} g^2(s)ds = 0$, we will have
that $H^{\ast}(\xi_{t_i},t_i)=F(Z_{t_i},t_i)$, therefore $H^{\ast}$
is an admissible and rational pricing rule. Moreover, if condition
(\ref{eq:cond1_nm}) is satisfied, then $\theta^{\ast}$ is continuous
with bounded variation. Therefore it follows from Proposition
\ref{char_tr_str_nm} that $\theta^*$ is optimal if condition
(\ref{eq:cond2_nm}) holds (notice that $H^*$ satisfies PDE
(\ref{price_HJB_nm})).

Thus, to show that $Y^{\ast}$ satisfies (\ref{eq:cond1_nm}) and
(\ref{eq:cond2_nm}) is the next goal. The proof is by induction.
\begin{description}
\item{I)} Consider the interval $[0,t_1]$. At $t=0$ we have $Y_0=0$, $Z_0=v$
and $Y_t$ satisfies the following stochastic differential equation
on $[0,t_1]$:
\begin{equation*}
dY_t=\frac{\left(Z_t-\alpha_{1}Y_t\right)\alpha_{1}}{\Sigma_z(t)+\sigma^2-\alpha_1^2
t }dt+dB^2_t
\end{equation*}
with $dZ_t=\sigma_z(t)dB^1_t$. From Theorem 10.3 of \cite{L} (note that due to  (\ref{eq:sigma_c3}), the
conditions of the theorem are satisfied), we have that on the filtration $\left( \mathcal{F}%
_{t}^{Y}\right) _{t< t_1}$ the stochastic differential equation for
$Y$ is
\begin{equation*}
dY_{s}=\frac{\left(m_{s}-\alpha_1
Y_{s}\right)\alpha_1}{\Sigma_z(t)+\sigma^2-\alpha_1^2 t
}ds+dB_{s}^{Y},
\end{equation*}
with
\begin{equation*}
dm_{s}=\frac{\gamma
_{s}\alpha_1}{\Sigma_z(t)+\sigma^2-\alpha_1^2t}dB_{s}^{Y},
\end{equation*}%
where $B_{t}^{Y}$ is Brownian motion on $\mathcal{F}_{t}^{Y}$, and
$\gamma _{s}$ satisfies the following ODE
\begin{equation*}
\dot{\gamma _{s}}=\sigma _{z}^{2}(s)-\frac{\gamma
_{s}^{2}\alpha_1^2}{\left( \Sigma_z(t)+\sigma^2-\alpha_1^2 t \right)
^{2}}
\end{equation*}%
with initial condition $\gamma _{0}=\sigma ^{2}$.

Notice that $\gamma _{s}=\Sigma_z(t)+\sigma^2-\alpha_1^2 t $ is the
unique
solution of this ODE and initial condition. Therefore on $\left( \mathcal{F}%
_{t}^{Y}\right) _{t\leq t_1}$, the process $Y$ satisfies
\begin{equation*}
dY_{s}=\frac{\left(B_{s}^{Y}-Y_{s}\right)\alpha_1^2}{\Sigma_z(t)+\sigma^2-\alpha_1^2
t }ds+dB_{s}^{Y}.
\end{equation*}%
The unique strong solution of this stochastic differential equation
on $[0,t_1) $ is $Y_{s}=B_{s}^{Y}$ (see \cite{Ka2},
Example 5.2.4). Hence, on the interval $[0,t_1)$, the process $Y$ is
a Brownian motion on its own (completed) filtration. By continuity
of $Y$, this process is a Brownian motion on $[0,t_1]$. To prove
that $\alpha_1Y_{t_1}=Z_{t_1}$, notice that
\begin{equation*}
\alpha_1Y_{t}^{\ast }=Z_{t}+\lambda (t)\left( -v+\int_{0}^{t}\frac{\alpha_1}{\lambda (s)}%
dB_{s}^{2}-\int_{0}^{t}\frac{\sigma _{z}(s)}{\lambda
(s)}dB_{s}^{1}\right)
\end{equation*}%
where $\lambda (t)=\exp \left\{
-\int_{0}^{t}\frac{\alpha_1}{\Sigma_z(t)+\sigma^2-\alpha_1^2
t}ds\right\} $.

Note that a random variable $\int_{0}^{t}\frac{\alpha_1}{\lambda (s)}%
dB_{s}^{2}-\int_{0}^{t}\frac{\sigma _{z}(s)}{\lambda (s)}dB_{s}^{1}$
is normally distributed with mean $0$ and variance
$\int_{0}^{t}\frac{\alpha_1^2+\sigma _{z}^{2}(s)}{\lambda
^{2}(s)}ds$. Therefore, due to condition (\ref{eq:sigma_c2}), if $
\lim_{t\rightarrow t_1}\int_{0}^{t}\frac{\alpha^2_1+\sigma _{z}^{2}(s)}{\lambda ^{2}(s)}%
ds<\infty $, then $\xi_{t_1}=\alpha_1Y_{t_1}=Z_{t_1}$.

On the other hand, if $\lim_{t\rightarrow
t_1}\int_{0}^{t}\frac{\alpha^2_1+\sigma _{z}^{2}(s)}{\lambda
^{2}(s)}ds=\infty $, consider the process
\begin{equation*}
X_{t}=\int_{0}^{t}\frac{\alpha_1}{\lambda
(s)}dB_{s}^{2}-\int_{0}^{t}\frac{\sigma _{z}(s)}{\lambda
(s)}dB_{s}^{1},
\end{equation*}%
and a change of time $\tau (t)$ given by
\begin{equation*}
\int_{0}^{\tau (t)}\frac{\alpha_1^2+\sigma _{z}^{2}(s)}{\lambda
^{2}(s)}ds=t.
\end{equation*}%
Then, $W_{s}=X_{\tau (s)}$ is a Brownian motion. Hence, we can use
the law of iterated logarithm to get
\begin{eqnarray*}
\limsup_{s\rightarrow \infty }\frac{W_{s}}{\sqrt{2s\log \log s}} &=&1 \\
\liminf_{s\rightarrow \infty }\frac{W_{s}}{\sqrt{2s\log \log s}}
&=&-1
\end{eqnarray*}%
or, in the original time,
\begin{eqnarray*}
\limsup_{t\rightarrow t_1 }\frac{X_{t}}{\sqrt{2\Xi (t)\log \log (\Xi (t)})%
} &=&1 \\
\liminf_{t\rightarrow t_1 }\frac{X_{t}}{\sqrt{2\Xi (t)\log \log (\Xi (t)})%
} &=&-1
\end{eqnarray*}%
where $\Xi (t)=\int_{0}^{t}\frac{\alpha_1^2+\sigma
_{z}^{2}(s)}{\lambda ^{2}(s)}ds$. Due to the conditions
(\ref{eq:sigma_c1})-(\ref{eq:sigma_c4}) in this case we have
\begin{equation*}
\lim_{t\rightarrow t_1}\lambda ^{2}(t)\Xi (t)\log \log \left( \Xi
(t)\right) =0,
\end{equation*}%
therefore it follows that $\lim_{t\rightarrow t_1}\lambda
(t)X_{t}=0$, thus $ \xi_{t_1}=\alpha_1 Y_{t_1}=Z_{t_1}$.

\item{II)} Suppose $\alpha_j(Y_{t_{j}}-Y_{t_{j-1}})=Z_{t_{j}}-Z_{t_{j-1}}$ for any $j\leq i$. Consider the interval $[t_i,t_{i+1}]$. At $t=t_i$ we have
$\xi_{t_i}=Z_{t_i}$,
and $\tilde{Y}_t=Y_t-Y_{t_i}$ satisfies the following stochastic
differential equation on $[t_i,t_{i+1}]$:
\begin{equation*}
d\tilde{Y}_t=\frac{\left(\tilde{Z}_t-\alpha_{i+1}\tilde{Y}_t\right)\alpha_{i+1}}{\Sigma_z(t)-\Sigma_z(t_i)-\alpha_{i+1}^2
(t-t_i) }dt+dB^2_t
\end{equation*}
with $\tilde{Z}_t=Z_t-Z_{t_i}$, thus
$d\tilde{Z}_t=\sigma_z(t)dB^1_t$ and $\tilde{Z}_{t_i}=0$. From
Theorem 10.3 of \cite{L} (note that, due to
(\ref{eq:sigma_c3}), the
conditions of the theorem are satisfied), we have that on the filtration $\left( \mathcal{F}%
_{t}^{Y}\right) _{t\in[t_i,t_{i+1})}$ the stochastic differential
equation for $Y$ is
\begin{equation*}
d\tilde{Y}_{s}=\frac{\left(m_{s}-\alpha_{i+1}
\tilde{Y}_{s}\right)\alpha_{i+1}}{\Sigma_z(t)-\Sigma_z(t_i)-\alpha_{i+1}^2
(t-t_i)}ds+dB_{s}^{Y},
\end{equation*}
with
\begin{equation*}
dm_{s}=\frac{\gamma
_{s}\alpha_{i+1}}{\Sigma_z(t)-\Sigma_z(t_i)-\alpha_{i+1}^2 (t-t_i)
}dB_{s}^{Y},
\end{equation*}%
where $B_{t}^{Y}$ is Brownian motion on $\mathcal{F}_{t}^{Y}$, and
$\gamma _{s}$ satisfies the following ODE
\begin{equation*}
\dot{\gamma _{s}}=\sigma _{z}^{2}(s)-\frac{\gamma
_{s}^{2}\alpha_{i+1}^2}{\left(
\Sigma_z(t)-\Sigma_z(t_i)-\alpha_{i+1}^2 (t-t_i) \right) ^{2}}
\end{equation*}%
with initial condition $\gamma _{t_i}=0$.

Notice that $\gamma _{s}=\Sigma_z(t)-\Sigma_z(t_i)-\alpha_{i+1}^2
(t-t_i)$ is the unique
solution of this ODE and initial condition. Therefore on $\left( \mathcal{F}%
_{t}^{Y}\right) _{t\in [t_i,t_{i+1})}$, the process $Y$ satisfies
\begin{equation*}
dY_{s}=\frac{\left(B_{s}^{Y}-Y_{s}\right)\alpha_{i+1}^2}{\Sigma_z(t)-\Sigma_z(t_i)-\alpha_{i+1}^2
(t-t_i) }ds+dB_{s}^{Y}.
\end{equation*}%
The unique strong solution of this stochastic differential equation
on $[t_i,t_{i+1})$ is $Y_{s}=B_{s}^{Y}$ (see \cite{Ka2}, Example 5.2.4). Hence, on the interval $[t_i,t_{i+1})$, the
process $Y$ is a Brownian motion on its own (completed) filtration.
By continuity of $Y$, this process is a Brownian motion on
$[t_i,t_{i+1}]$. To prove that
$\alpha_{i+1}\tilde{Y}_{t_{i+1}}=\tilde{Z}_{t_{i+1}}$, notice that
\begin{equation*}
\alpha_{i+1}Y_{t}^{\ast }=Z_{t}+\lambda (t)\left(\int_{t_i}^{t}\frac{\alpha_{i+1}}{\lambda (s)}%
dB_{s}^{2}-\int_{t_i}^{t}\frac{\sigma _{z}(s)}{\lambda
(s)}dB_{s}^{1}\right)
\end{equation*}%
where $\lambda (t)=\exp \left\{
-\int_{t_i}^{t}\frac{\alpha_{i+1}}{\Sigma_z(t)-\Sigma_z(t_i)-\alpha_{i+1}^2
(t-t_i) }ds\right\} $.

Therefore, due to conditions (\ref{eq:sigma_c1})-(\ref{eq:sigma_c4})
we have, exactly as in the previous case,
$\alpha_{i+1}\tilde{Y}_{t_{i+1}}=\tilde{Z}_{t_{i+1}}$.
\end{description}
By the principle of mathematical induction, conditions
(\ref{eq:cond1_nm}) and (\ref{eq:cond2_nm}) hold for any $i$.
\end{proof}
\section{Conclusion}

This paper demonstrates that, in the presence of \emph{dynamic}
private information of the insider, and under minimal restrictions
on the admissible trading strategies, an equilibrium exists and
there is a unique Markovian pricing rule (as a function of total
order process) that admits inconspicuous equilibrium. Moreover, the
optimal insider trading strategy is based on the market estimates of
the fundamentals, rather than on the stock price: the insider buys
the stock when the market overestimates the fundamental value, and
sells it otherwise, thus leading to higher informativeness of the
stock price. Furthermore, this induces convergence of the price to
the fundamental value at the terminal time in the case of Markovian
pricing rule and at some some set of times (which include the terminal
time) in the case of non Markovian pricing rule.

Future research can be conducted along the following directions:
assumptions on the market parameters could be further relaxed, and a
more general model of the total order of the noisy traders could be
considered. The model can also be generalized further by allowing
for potential bankruptcy of the firm issuing the stock, with the
time of bankruptcy defined as the random time at which the
underlying process governing the firm value hits a given barrier.

\newpage
\appendix
\section{Proof of Remark {\ref{speed_of_conv}}}
Suppose Assumption \ref{ass2} is satisfied, $\lim_{t\rightarrow
1}\Xi(t)=\infty$ and conditions (\ref{speed_of_adj_infty}) and
(\ref{eq:signal_precision}) hold. Then L'H\^{o}pital rule will give
(notice that due to (\ref{eq:signal_precision}) and continuity of $\sigma_z(t)$ in the vicinity of $1$ we have $\lim_{t\rightarrow
1}\left(1+\sigma^2_z(t)\right)<2$)
\begin{equation*}
\lim_{t\rightarrow
1}\lambda^2(t)\Xi(t)\log\log\left(\Xi(t)\right)=\frac{1}{2}\lim_{t\rightarrow
1}\left(1+\sigma^2_z(t)\right)\lim_{t\rightarrow
1}\left(\Sigma_z(t)+\sigma^2-t\right)\log\log\left(\Xi(t)\right).
\end{equation*}
Since by L'H\^{o}pital rule we have
\begin{equation*}
\lim_{t\rightarrow 1}\lambda^2(t)\Xi(t)=0,
\end{equation*}
 it follows that
 \begin{eqnarray*}
0 &\leq & \lim_{t\rightarrow
1}\lambda^2(t)\Xi(t)\log\log\left(\Xi(t)\right)\leq
\frac{1}{2}\lim_{t\rightarrow
1}\left(1+\sigma^2_z(t)\right)\lim_{t\rightarrow
1}\left(\Sigma_z(t)+\sigma^2-t\right)\log\log\left(\lambda^{-2}(t)\right)\\
&=&\frac{1}{2}\lim_{t\rightarrow
1}\left(1+\sigma^2_z(t)\right)\lim_{t\rightarrow
1}\left(\Sigma_z(t)+\sigma^2-t\right)\log\left(\int_0^t\frac{1}{\Sigma_z(s)+\sigma^2-s}ds\right)\\
 &=& \frac{1}{2}\lim_{t\rightarrow
1}\left(1+\sigma^2_z(t)\right)\lim_{t\rightarrow
1}\frac{\log(f(t))}{f'(t)},
\end{eqnarray*}
where $f(t)=\int_0^t\frac{1}{\Sigma_z(s)+\sigma^2-s}ds$ and $\lim_{t\rightarrow
1}f(t)=\infty$.
Since $\lim_{x\rightarrow\infty}\frac{\log\left(x\right)}{x^{\alpha}}=0$, for any $\alpha>0$
we need to show that
\begin{equation}\label{eq:app1}
\limsup_{t\rightarrow
1}\frac{f^{\alpha}(t)}{f'(t)}<\infty
\end{equation}
for some $\alpha>0$ to establish (\ref{xi_cond}).

Consider any $\alpha\in(0,1)$ and denote by
\begin{equation*}
0<g(t)=\frac{f^{\alpha}(t)}{f'(t)},
\end{equation*}
then for $t\geq t^*$ we have
\begin{equation*}
f^{1-\alpha}(t)=(1-\alpha)\int_{t^*}^t\frac{1}{g(s)}ds+c
\end{equation*}
 where $c$ is some positive constant. Due to this expression and since $\lim_{t\rightarrow 1} f(t)=\infty$, $\alpha<1$ and, due to (\ref{speed_of_adj}) , $f(t)<\infty$ for any $t\in [0,1)$  we must have
\begin{equation*}
\lim_{t\rightarrow
1}g(t)=0.
\end{equation*}
Thus (\ref{eq:app1}) holds and therefore (\ref{xi_cond}) is
established.

\end{document}